\documentclass[prb,singlecolumn,showpacs,preprintnumbers,amsmath,amssymb]{revtex4}

\usepackage{amsfonts}
\usepackage{latexsym}
\usepackage{graphicx}
\usepackage[usenames]{color}
\usepackage[utf8]{inputenc}
\usepackage[english]{babel}
\usepackage{hyperref}

\begin{document}
\title{Resonance measurement of periodically driven contact potential difference}
\author{Vasil G. Yordanov and Todor M. Mishonov}

\email[E-mail: ]{vasil.yordanov@gmail.com, mishonov@gmail.com}
\affiliation{Department of Theoretical Physics, Faculty of Physics,\\
Sofia University St. Clement of Ohrid,\\
 5 James Bourchier Blvd., BG-1164 Sofia, Bulgaria}

 
\date{February 25, 2015}

\begin{abstract}
A new type of quartz resonance device for measurements of oscillations of the contact potential difference induced by modulated light is described. 
Special attention is devoted to the compensation of the constructive capacitance of the quartz resonator by a negative capacitance. 
In such a way the the quartz resonance filter is very narrow and at the same time has extremely small total bandwidth, which improves significantly the signal to noise ratio. 
For metals, this device gives the temperature dependence of the work function and opens perspectives for creation of imaging spectroscopy based on this temperature derivative. 
The proposed device can be used for creation of defectoscopy of metallic materials based on the temperature derivative of the work function.  
\end{abstract}

\maketitle

\section{Introduction}
In the classical vibrating reed method by Kelivin~\cite{Kelvin1898} the AC current pass through a high Ohmic resistance and the corresponding AC voltage, on the resistor ends, determines the constant Contact Potential Difference (CPD). For a contemporary application of this method see, for example, the PhD thesis.\cite{Lagel2000}

If the CPD is not constant in time, and it is induced by chopping of the laser radiation or by some other electromagnetic excitations, then the classical vibrating reed method is not applicable and we suggest to use a resonance method for measurement of this Periodically Driver CPD (PD-CPD).
We describe a simple electronic scheme for resonance measurements of PD-CPD, which can find many applications from the 
contactless control of raw materials~\cite{Ivanov2012} and milk quality control~\cite{Ivanov2009} to the fundamental properties of modern materials in condensed mater physics, such as superconductors~\cite{Mishonov1994a} and semiconductors.~\cite{Das1992} 
For metals, oscillation of the temperature by chopped laser heating can give directly temperature derivative of CPD.
This derivative is directly related to the logarithmic derivative of the density of states at the Fermi level. 
For quasi two dimensional conductors the CPD oscillations change the sign when the topology of the Fermi contour is changed between hole and electron one~\cite{Mishonov2006}.

The work is organized as follows. In the next Section~\ref{sec:classical_CPD} the classical method for measurement of CPD is described. After that in Section~\ref{sec:resonance_CPD} the theory of resonance determination of PD-CPD is given. In this chapter the equivalent circuit of quartz resonator and the concept of negative capacitance used to improve the work of the quartz resonator is presented. 
In Section~\ref{sec:practical_impl} a practical implementation of resonance measurement of PD-CPD can be found. Finally in Section~\ref{sec:transfer_func} the transfer function of the proposed electric circuit for measurement of periodically driven CPD is given.

%
\section{Classical Contact Potential Difference measurements}
\label{sec:classical_CPD}
\begin{figure}[ht]
  \includegraphics[width=.8\linewidth]{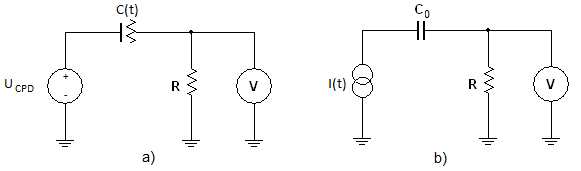}
  \caption{Classical scheme of measuring CPD by vibrating reed capacitor $C(t)$. Voltage drop on the resistor $R$ determines CPD.
a) Time dependent capacitance $C(t)$ and constant $U_\mathrm{CPD}$ create current $I(t)=\dot{C}(t) U_\mathrm{CPD}$.
b) The equivalent scheme is a current generator which creates current $I(t)$ through the resistor $R$ and constant capacitor $C_0$.
The voltage drop on the resistor is $U(t)=(\tau \dot{C}(t)/C_0) U_\mathrm{CPD}$.}
  \label{fig:Classical_circuit_U}
\end{figure}

Classical scheme of measuring CPD~\cite{Kelvin1898,Zisman1932,Lagel2000} is given in FIG.~\ref{fig:Classical_circuit_U}.
The vibrating reed creates time dependent capactance $C(t)$, which is a periodic function $C(t)=\epsilon_0 S/(d_0+A \cos(\omega t))$, where $S$ is the surface of the vibrating reed and $A$ is the amplitude of the vibrations around the mean distance between plates $d_0$. 
The resistance $R$ has usually high ohmic value. For example, in the classical experiment by Zisman~\cite{Zisman1932} $R=20\, \mathrm{M}\Omega$. 

The CPD creates a charge on the capacitor $Q(t)=C(t) U_\mathrm{CPD}$, see FIG.~\ref{fig:Classical_circuit_U}. 
For superconductors $C$ could be constant and the currents can create current induced CPD.~\cite{Mishonov1994a} The same could be for Field Effect Transistors (FET).~\cite{Das1992}
Source-Drain currents can create additional polarization of gate charge.
However for the classical setup, CPD is constant and the time dependence of the charge is generated by vibrating reed, which creates time dependent capacitance $C(t)$.

For low frequencies of the vibrations current through the resistor $I(t)=\mathrm{d}_t(C U_\mathrm{CPD})=\dot{C} U_\mathrm{CPD}$ in the effective scheme depicted at FIG.~\ref{fig:Classical_circuit_U}
creates a voltage drop on the resistor 
\begin{equation}
\label{eq:U_CPD}
U(t)=R \dot{Q}(t)=R I(t)=R \dot{C}(t)\, U_\mathrm{CPD} = \frac{\tau\dot{C}(t)}{C_0}\, U_\mathrm{CPD},
\end{equation}
where $\tau=R C_0$, $C_0$ is the non-perturbed by vibrations capacitance and dot and $\mathrm{d}_t$ denote time differentiation.
The quasi-static formula for the current through the capacitor $I(t)= \dot{C} U_\mathrm{CPD}$ is applicable for frequencies $f \ll 1/(2 \pi R C_0)$. 

In the next section we will describe a modification of the method when the capacitor is not vibrating, but the CPD is driven by periodical excitations.

\section{Theory of resonance determination of periodically driven CPD}
\label{sec:resonance_CPD}
For time dependent CPD $U_\mathrm{CPD}(t)$ the current $I(t)$ according to FIG.~\ref{fig:Classical_circuit_U}b is again time derivative of the charge, but the charge is determined by the time dependent CPD.
In this case the voltage drop on the resistor $R$ is  
\begin{equation}
\label{eq:U_PD_CPD}
U(t)=\tau\dot{U}_\mathrm{CPD}(t), \quad \tau=R C_0.
\end{equation}
As though in FIG.~\ref{fig:Classical_circuit_U}b the current source is replaced by voltage source $U_\mathrm{CPD}(t)$.
This effective voltage generator is hidden by the capacitive connection to the sample and our task is to measure its small oscillations.
We emphasize ``effective'', because if the capacitor is removed then the effect disappears. CPD can not be measured by voltmeter.
Even Volta has constructed electrometer and succeed to measure mV CPD.

\begin{figure}[ht]
\includegraphics[scale=0.8]{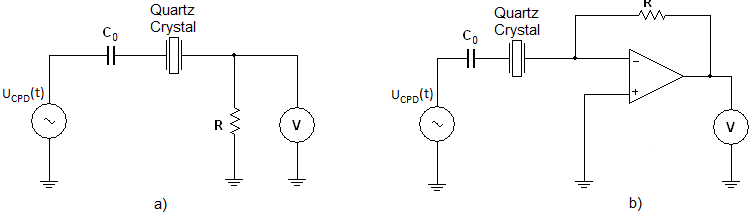}
\caption{Resonance scheme for measuring of time dependent CPD with quartz resonator in series with the capacitive connection to the sample.}
\label{general_CPD_X_circuit}
\end{figure}

\begin{figure}[ht]
\includegraphics[scale=0.8]{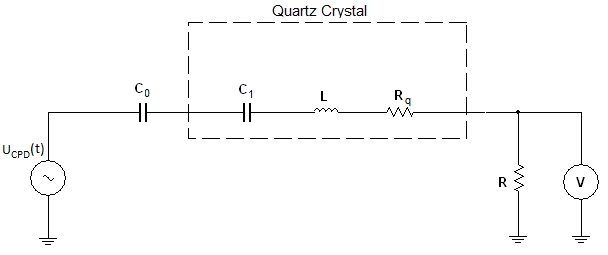}
\caption{Simplified equivalent schematic of quartz resonator, with ignored parasitic capacitance $C_2$, in the electric circuit for measuring PD-CPD.
The capacitive connection $C_0$ is in series with the small capacitance $C_1$ of the quartz resonator. }
\label{general_CPD_X_equivalent_circuit_ignored_C0}
\end{figure}

We propose a method for measurement of time dependent CPD, when the CPD oscillations are periodically excited 
\begin{equation}
U_\mathrm{PD-CPD}(t)=\Re \left( \hat{U}_\mathrm{PD-CPD}\,\mathrm{e}^{\mathrm{j}\omega t} \right),
\end{equation}
where j=-i. 
In this method a quartz resonator is connected in series with a capacitive connection $C_0$ to the source of the signal -- time dependent CPD, see FIG.~\ref{general_CPD_X_circuit}. 
The capacitive connection to the sample operates as a load capacitance of the quartz resonator, changing a little bit its resonance frequency. 
At the resonance frequency the quartz resonator in the first approximation is real resistor $R_\mathrm{q}$, see FIG.~\ref{general_CPD_X_equivalent_circuit_ignored_C0}.
The voltage drop on $R$ from FIG.~\ref{general_CPD_X_circuit}a and FIG.~\ref{general_CPD_X_equivalent_circuit_ignored_C0} is 
\begin{equation}
\label{eq:U_PD_CPD}
U(t) \approx \frac{R}{R+R_\mathrm{q}}\, U_\mathrm{PD-CPD}(t).
\end{equation}
In resonance the impedance of inductance $L$ and capacitors $C_0$ and $C_1$ are canceled and the quartz resonator is like a real ohmic resistor.

Alternative way to measure the $U_\mathrm{PD-CPD}(t)$ is to use a current to voltage converter as it is shown in FIG.~\ref{general_CPD_X_circuit}b.
In this case the output voltage at the resonance frequency is given by the formula of an inverting amplifier
\begin{equation}
\label{eq:U_PD_CPD}
U(t) \approx -\frac{R_\mathrm{f}}{R_\mathrm{q}}\, U_\mathrm{PD-CPD}(t).
\end{equation}
At the analysis of the thermal noise of the electronic schemes from FIG.~\ref{general_CPD_X_circuit}a and FIG.~\ref{general_CPD_X_circuit}b the resistor $R$ respectively $R_\mathrm{f}$ is parallel to the effective resistor $R_\mathrm{q}$ of the quartz resonator at the resonance frequency. That is why for $R,\, R_\mathrm{f} \gg R_\mathrm{q}$ the signal to noise ration is determined by the thermal noise of the quartz resonator.

\begin{figure}[ht]
\includegraphics[scale=0.8]{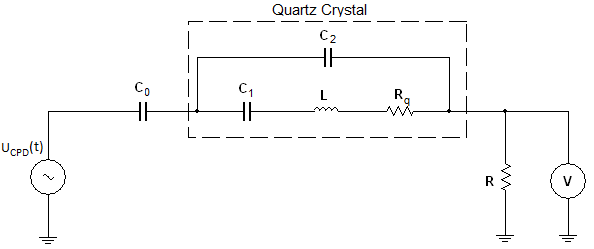}
\caption{Equivalent Van Dyke scheme~\cite{Dyke1928} of quartz resonator in the electric circuit for measuring PD-CPD.}
\label{general_CPD_X_equivalent_circuit}
\end{figure}

Every quartz resonator has a constructive capacitance in the effective scheme by Van Dyke~\cite{Dyke1928} given in FIG.~\ref{general_CPD_X_equivalent_circuit}. 

In this figure $C_2$ is a constructive capacitance formed by the plates surrounding the quartz crystal. This capacitance is of the order of few pF.
$C_1$ is the capacitance due to the elastic properties of the resonator. For the used in this work clock quartz resonator the capacitance is in the order of few fF. 
$L$ is the inductance of the quartz resonator due to its mass. 
$R_\mathrm{q}$ is the ohmic resistance of quartz resonator due to mechanic dissipations. 
The $Q$ factor of the clock quartz resonators is in range of $10^4 - 10^5$.
On the scheme in FIG.~\ref{general_CPD_X_equivalent_circuit} at the series resonance the sequential impedances of $C_1$ and $L$ are canceled  and impedance $R_\mathrm{q}$ is much smaller than the modulus of impedance of the constructive capacitance $C_2$, i.e. $R_\mathrm{q} C_2 \omega_\mathrm{res} \ll 1$.
As though the capacitor $C_2$ disappears and the capacitor $C_0$ becomes in series with the capacitor $C_1$. 
If we take into account that $C_1 \ll C_0$ then small changes of the capacitance $C_0$ of the capacitive connection will cause small changes of the resonant frequency.
On the other hand the ration of complex amplitudes between the voltage drop on $R$ and $U_\mathrm{CPD}$ depends on $C_0$
\begin{equation}
\Upsilon=\frac{\hat{U}}{\hat{U}_\mathrm{CPD}}\approx\frac{1}{1/(\mathrm{j}\, \omega_\mathrm{res} R C_0)+(R_\mathrm{q}/R)+1}, 
\end{equation}
where $\omega_\mathrm{res}$ is the resonance frequency.
That is why in the present method it is important to keep the capacitance of the connection to the sample to be constant and not to change during the measurements.

\begin{figure}[ht]
\includegraphics[scale=0.7]{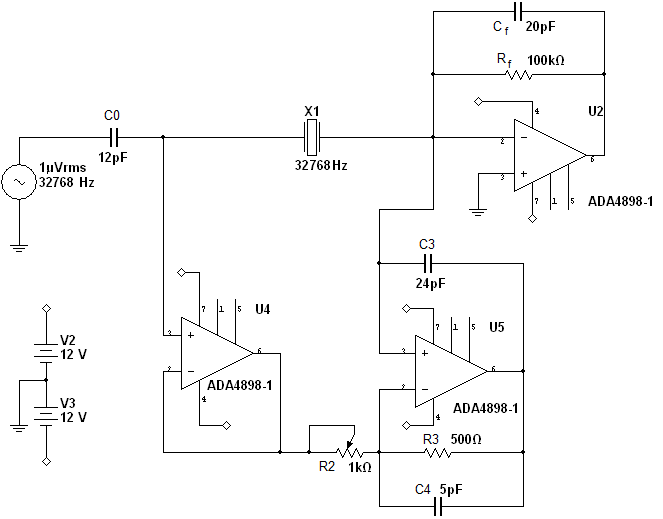}
\caption{Resonance amplifier with compensated constructive capacitance. This scheme works as an inverting amplifier with big value of $R_\mathrm{f}$ and high Q sequential resonance LRC, scheme represented in FIG.~\ref{general_CPD_X_equivalent_circuit_ignored_C0}, switched to the inverting input of OpAmp.
Right down corner with OpAmp U5 works as a negative capacitor.}
\label{modified_ivanov_schema}
\end{figure}

The constructive capacitance $C_2$ in FIG.~\ref{general_CPD_X_equivalent_circuit} gives a short cut for the high frequency signals which significantly decrease the quality of the resonator.
In order to compensate this parasitic capacitance it is useful to connect a negative capacitance $-C_2$ in parallel to the quartz resonator.
The negative capacitance can be implemented by the well known scheme of the negative impedances.~\cite{Gordon1970} 
This simple idea is realized in the scheme depicted in FIG.~\ref{modified_ivanov_schema}. 
In such a way we obtain the sequential resonator depicted on FIG.~\ref{general_CPD_X_equivalent_circuit_ignored_C0}.
To the best of our knowledge for the first time compensation of parasitic capacitance of Van Dyke scheme was presented in the work by Ivanov.\cite{Ivanov1999}.
Here we describe an alternative realization of this simple idea.
The first Operational Amplifier (OpAmp) U4 is voltage follower, which repeats the voltage applied to the quartz resonator. 
The OpAmp U5 together with R2, R3, C3 implements a negative capacitor, which exactly compensate the capacitor $C_2$ if $C_2 R_2 = C_3 R_3$. 
To satisfy this condition one can trim the variable resistor $R_2$.
The additional capacitor C4 dumps the parasitic oscillations. 
For low frequency the negative input of U2 is virtual ground.
In this virtual ground the quartz and the output of the negative capacitor are switched.
This gives the compensation of the constructive capacitor of quartz crystal in the scheme of Van Dyke FIG.~\ref{general_CPD_X_equivalent_circuit}.
U2 operates as and inverting amplifier, which at the resonance has amplification 
\begin{equation}
\Upsilon=-\frac{R_\mathrm{f}}{R_\mathrm{q}},
\end{equation}
for non-resonance frequencies the amplification is $|\Upsilon| \ll 1$. 
More precisely 
\begin{equation}
\Upsilon(\omega)=-\frac{R_\mathrm{f}/(1+\mathrm{j}\omega R_\mathrm{f}C_\mathrm{f})}{R_\mathrm{q}+\mathrm{j}\left[ \omega L - (C_0+C_1)/(C_1 C_0\omega) \right]}.
\end{equation}

Another non-ideality of the quartz resonator that we should take into account is that its resonance frequency depends on the environment temperature.
That is why it is necessary for the resonator to be mounted in a crystal oven -- temperature-controlled container. 
Or to use quartz resonator that has the maximum of its frequency at room temperature.

Benefits of the method for measurement of PD-CPD presented in this work are:
small impedance in resonance -- from several hundred ohms up to several tens of $\mathrm{k}\Omega$;
very narrow bandwidth due to the high Q factor;
small influences from external electromagnetic interferences;
good signal to noise ration due to the low thermal noise of the quartz resonator in resonance and due to the narrow bandwidth.

\section{Practical implementation of resonance measurement of PD-CPD}
\label{sec:practical_impl}
In this work we describe an implementation of electric circuit for measurement of PD-CPD.
In the next section we will show some experimental results obtained using the suggested circuit with simulated PD-CPD signal.

The main requirement is to create an electronics capable to amplify the signal up to 80 dB at 32 kHz, and in the same time to maintain low noise level so that the output signal can be measured by ordinary DC voltmeter. 
Another requirement for the electronics is to have small size and to be portable.
The implemented circuit is put in a small metal box with dimensions 3x10x12 cm, powered by two batteries of 9 V.

\begin{figure}[ht]
\includegraphics[scale=0.6]{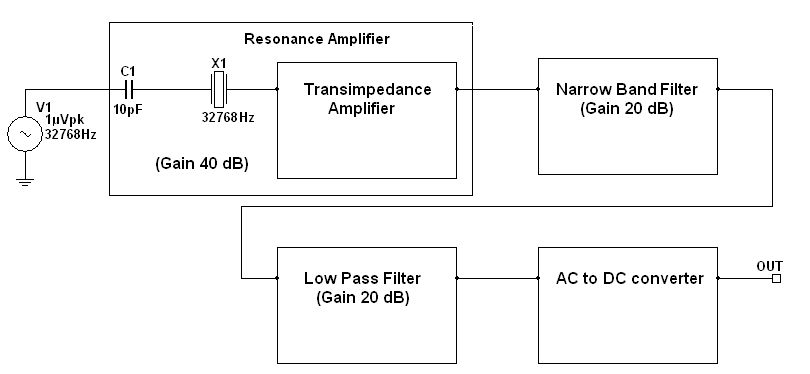}
\caption{Block scheme for resonance measurement of PD-CPD. The output voltage is measured directly with DC voltmeter.}
\label{block-diagram}
\end{figure}

\begin{figure}
\includegraphics[scale=0.5]{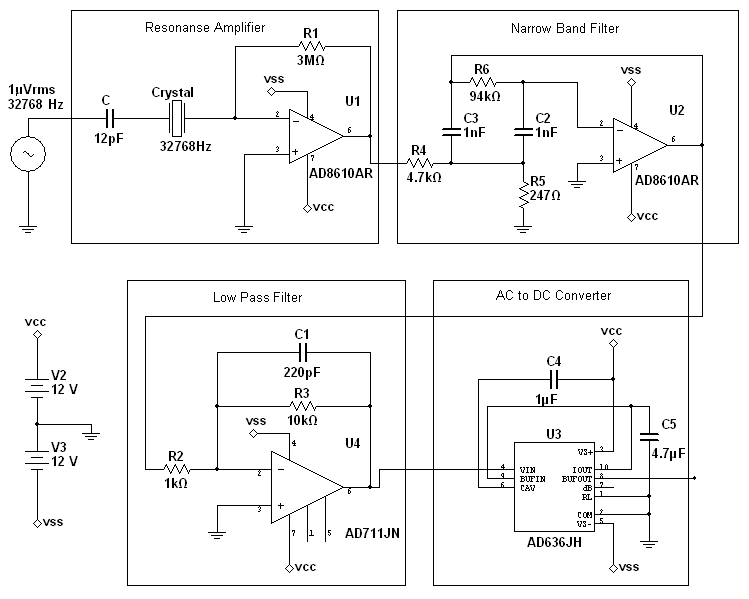}
\caption{Scheme for resonance measurement of CPD.}
\label{electronic_scheme}
\end{figure}

Block diagram of the implemented electric circuit is given in  FIG.~\ref{block-diagram}.
A detailed implementation of each block is presented in FIG.~\ref{electronic_scheme}.

The first amplifier is a transimpedance amplifier, which is a current to voltage converter, giving Gain 40 dB. The connection of this amplifier with the sample (source of the signal) as well as the principle of the operation was described in details in the previous section~\ref{sec:resonance_CPD} and FIG.~\ref{general_CPD_X_circuit}b. 
In the proposed circuit we have used a clock quartz resonator with resonance frequency of 32768Hz.
The impedance of the quartz in resonance is around $R_\mathrm{q}$=30 k$\Omega$, when the load capacitance $C_0=12$ pF. 
The noise analyze of the circuit gives that the quartz resonator is the main source of (thermal) noise, which is around 25 $\mathrm{nV}/\sqrt{\mathrm{Hz}}$ at room temperature. Another source of noise is the operational amplifier AD8610 with voltage noise about 6 $\mathrm{nV}/\sqrt{\mathrm{Hz}}$, which is much smaller than the noise of the quartz resonator in resonance. 
The current noise 5 $\mathrm{fA}/\sqrt{\mathrm{Hz}}$, of the used operation amplifier is negligible, due to the low input impedance of quartz resonator in resonance.
Gain Bandwidth Product (GBP) of AD8610 is around 25 MHz,~\cite{AD8610} which is not reached in the current circuit, because the chosen frequency is 32 kHz and the gain of the amplitude at resonance is 100.

\begin{figure}[ht]
\includegraphics[scale=0.5]{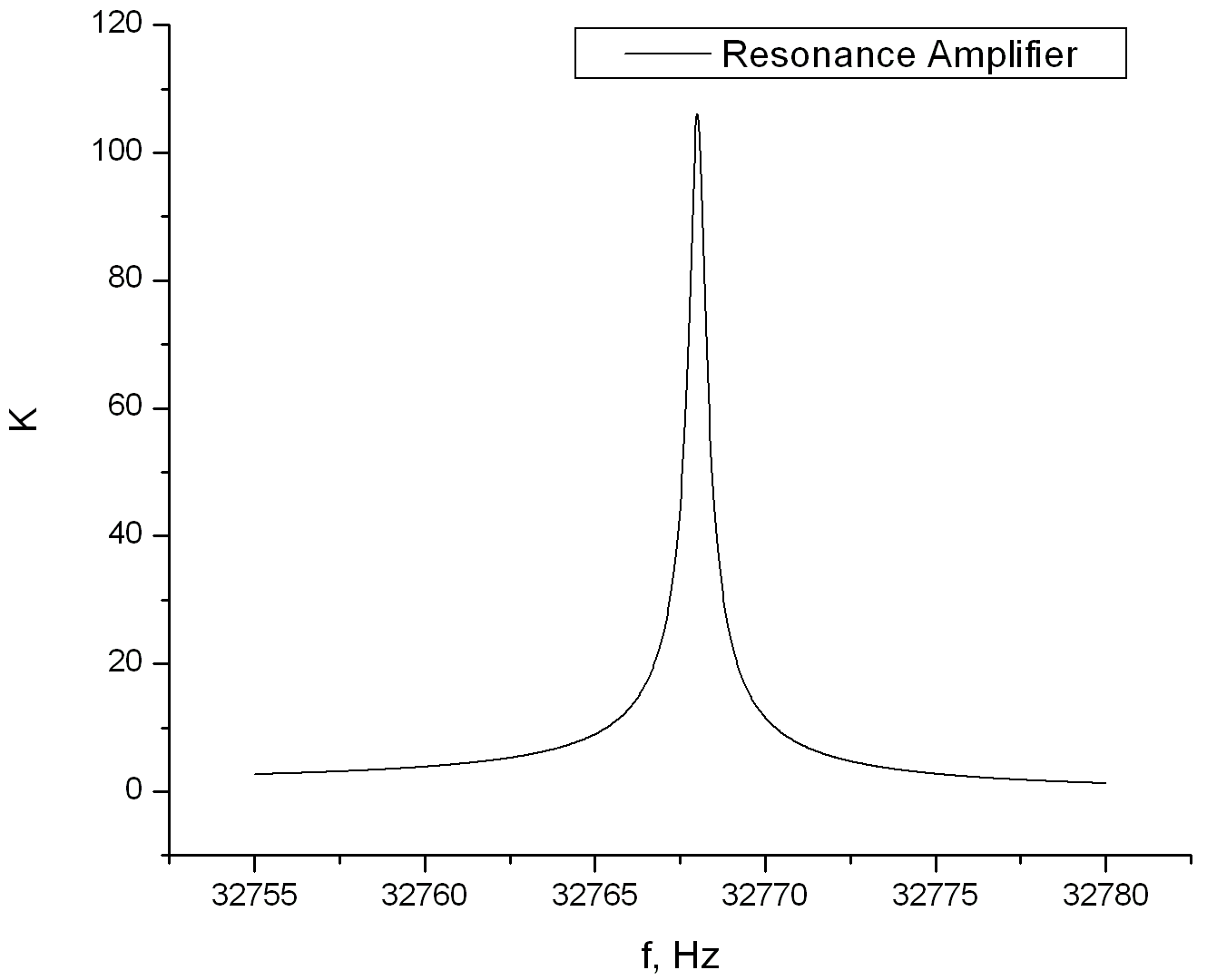}
\caption{Resonance amplifier with compensated shunt capacitance.}
\label{1st_module_k_w}
\end{figure}

Simulation Program with Integrated Circuit Emphasis (SPICE)~\cite{Nagel1973} simulation of the transfer function of this module is given in FIG.~\ref{1st_module_k_w}.
According to this figure the bandwidth of the resonance amplifier is approximately 1 Hz. Using this bandwidth and the noise spectral density we calculate the Root Mean Square (RMS) of the total noise to be around 25 nV.
In current implementation the amplification is 80 dB, due to the requirement that the entire circuit has to be placed in a single small metal box. That is why the total noise is amplified to 250 $\mu$V.

Next modules amplifies the signal and do some additional filtering by narrowing the band pass around the frequency of the measured signal.

\begin{figure}[ht]
\includegraphics[scale=0.5]{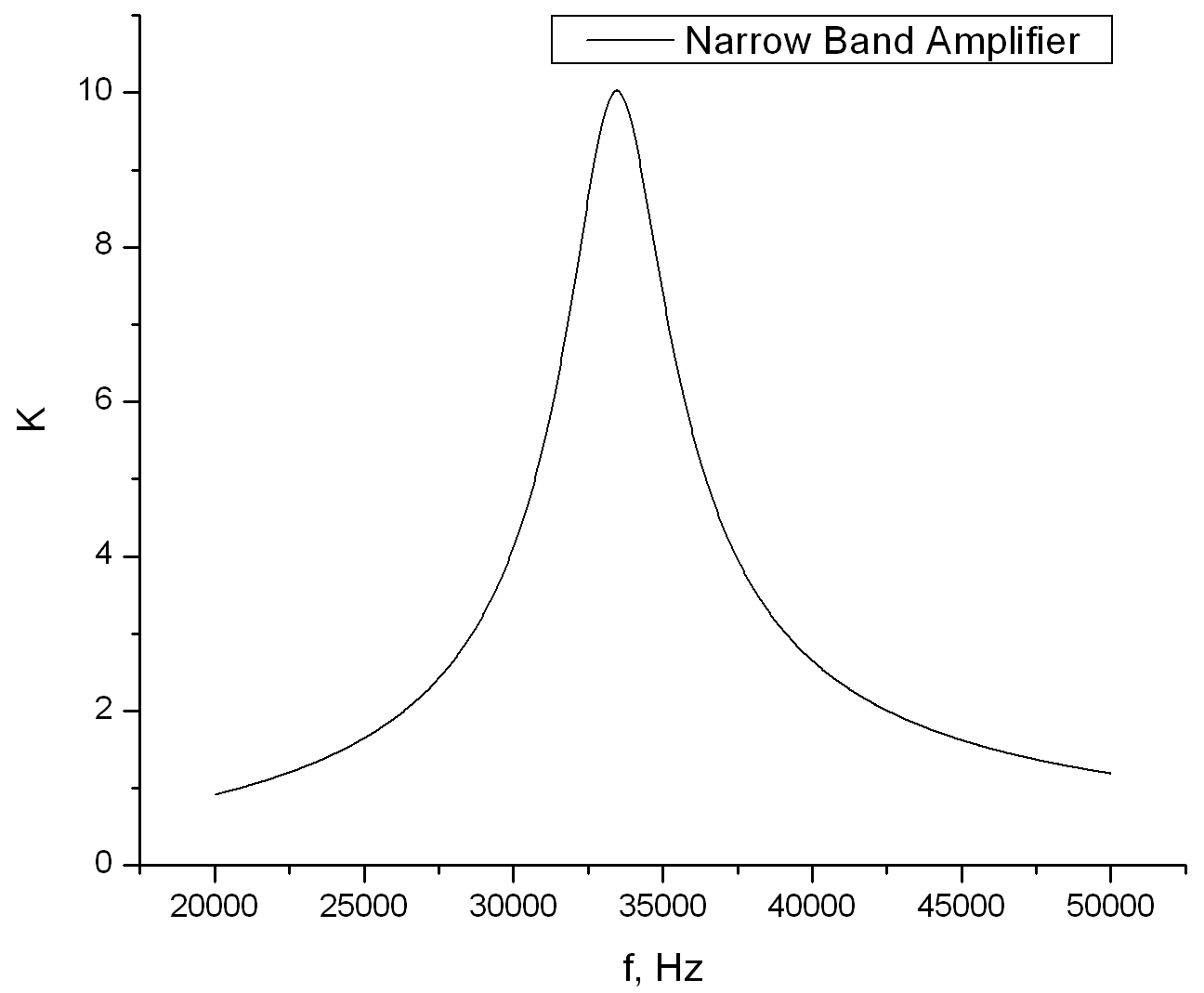}
\caption{Resonance amplifier with compensated shunt capacitance.}
\label{2nd_module_k_w}
\end{figure}

The second module is a narrow band filter with Gain 20 at the center frequency. This module is implemented according,~\cite{SLOA093} but any other similar circuit could be used.
SPICE simulation of the transfer function of this filter is given in FIG.~\ref{2nd_module_k_w}. 
From this simulation one can see that the bandwidth of the narrow band filter is much wider than the bandwidth of the first resonance module implemented with quartz resonator. 
The benefit of using wider filter in the second module is that it is easy to adjust its center frequency to the frequency of the first resonance amplifier. 
Another benefit of this filter is the better filtering of the lower and the higher frequencies simultaneously, due to the high bandwidth of the used operational amplifier, i.e. it decrease the influence of the low- and high-frequency wings of the quartz resonator.
This is the way to achieve small total bandwidth.

The last AC module in FIG.~\ref{block-diagram} and FIG.~\ref{electronic_scheme} is a low pass filter with Gain 20. It is used to filter any frequencies higher than the frequency of the measured signal that have passed through the first two modules.
The filter is implemented using first order active low-pass filter schematics. 

After the first three AC modules we have implemented an AC to DC converter using the true RMS-to-DC converter AD636 from Analog Devices.~\cite{AD636} 
The used averaging time constant $T$ of the converter is equal to 120 ms. 
With this converter we are able to measure the amplitude of the PD-CPD directly with a DC voltmeter
\begin{equation}
\label{AC_DC}
U_\mathrm{DC}=\sqrt{\lim_{T \rightarrow \infty} \int_0^T U_\mathrm{AC}^2 (t) \frac{\mathrm{d}t}{T}}, 
\qquad \frac{1}{T}\ll f\ll f_\mathrm{cut-off},
\end{equation}
where we suppose that all typical frequencies of the signal are lower than cut-off frequencies of the AC-DC converter AD636, see figure~5 of Ref.~\onlinecite{AD636}, but greater than reciprocal time-averaging constant. Cut-off frequency amplitude dependent and for 100~kHz amplification should ensure $U_\mathrm{DC}>10$~mV.

\section{Transfer function of the proposed electric circuit for measurement of periodically driven CPD}
\label{sec:transfer_func}
In this section we will present experimental results from measurement of transfer function of the electric circuit proposed in the previous chapter. 

The output voltage is measured with a DC voltmeter UNI-T 60A.~\cite{UT60A}
The digital generator RIGOL-DG1022~\cite{DG1022} is used as an AC voltage source generator. 
The output of the generator is attenuated 1000 times by a resistor divider with resistors 10 k$\Omega$ and 10~$\Omega$, giving output voltage from 4~$\mu$V up to 600~$\mu$V.

\begin{figure}[ht]
\includegraphics[scale=0.5]{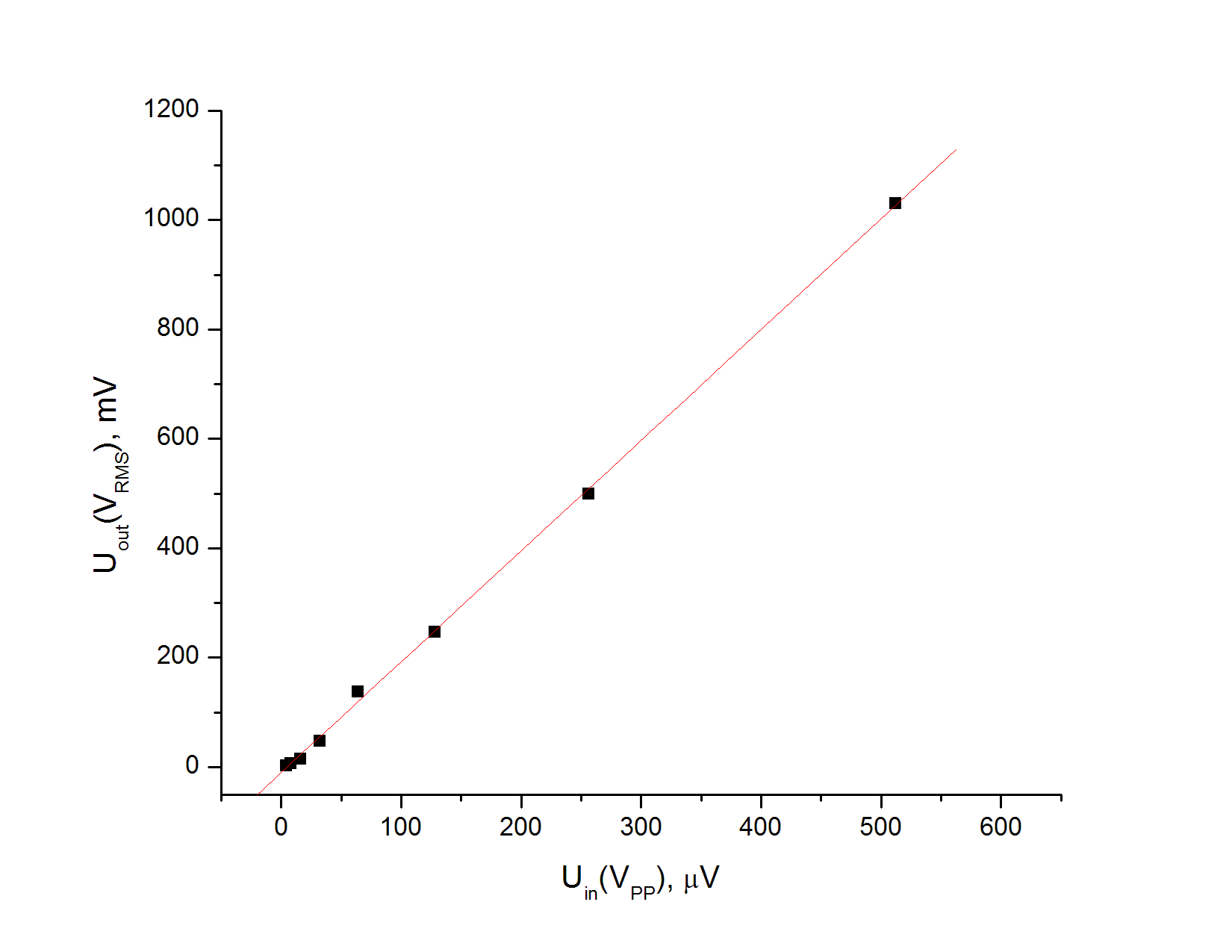}
\caption{Resonance amplifier with compensated shunt capacitance.}
\label{SelectiveAmlifierResponse}
\end{figure}

Results from the measurement 	are presented in FIG.~\ref{SelectiveAmlifierResponse}. From this figure it is seen that transfer function is linear for input voltages above 100 $\mu$V and has small deviations from the linearity for voltages below 100 $\mu$V. 
The nonlinearity in the transfer function at small input signal levels is mainly due to the nonlinearities in the AC-DC converter AD636.~\cite{AD636}

The smallest voltage signal measured with this electronics is 4 $\mu$V, which is much higher than the total noise referenced to the input of the amplifier 25 nV. 
This means that there is more room for AC voltage amplification, but some special considerations related to the shielding of the circuit should be taken into account when implementing the AC amplifier with higher amplification, especially if this amplifier is put into a single metal box.
For example at signal to noise ration equal to 10 this circuit is capable of measuring a signal around 250 nV, without any lock-in amplification if there is enough AC voltage amplification.

\section{Discussion and conclusion}
The method suggested by us for measuring periodically driven CPD can be applied for fundamental research in condensed matter physics, for example, the observation of nano-Volt current induced CPD in superconductors.~\cite{Mishonov1994a}
Or for solution of some practical problems as ``surface photon-charge effect in conductors'' Refs.~\onlinecite{Ivanov1989,Ivanov1990}.
For metals the described device can measure the periodic modulation of Fermi level induced by surface temperature modulation using chopped light or modulated laser power. 
After some additional analysis the measurement can give temperature derivative of the work function of the metal $\mathrm{d}W/\mathrm{d}T$.

\section*{Acknowledgment}
The authors are thankful to Mario Metodiev for the critical reading of manuscript.


\end{document}